\documentclass[twocolumn,prd,nofootinbib]{revtex4}

\newif\ifpdf
\ifx\pdfoutput\undefined
\pdffalse 
\else
\pdfoutput=1 
\pdftrue
\fi

\ifpdf
\usepackage[pdftex]{graphicx}
\else
\usepackage{graphicx}
\fi

\usepackage{epsf}
\usepackage{dcolumn}

\newcommand{\be}{\begin{equation}}
\newcommand{\ee}{\end{equation}}

\def\ie{{\it i.e.},~}
\def\eg{{\it e.g.},~}

\def\4he{$^4$He}
\def\3he{$^3$He}
\def\7li{$^7$Li}

\newcommand\la{\lower0.6ex\vbox{\hbox{\ensuremath{\buildrel{\textstyle<}\over{\sim}\ }}}}
\newcommand\ga{\lower0.6ex\vbox{\hbox{\ensuremath{\buildrel{\textstyle>}\over{\sim}\ }}}}

\def\vev#1{{\langle#1\rangle}}

\def\lsim{\mathrel{\raise.3ex\hbox{$<$\kern-.75em\lower1ex\hbox{$\sim$}}}}
\def\gsim{\mathrel{\raise.3ex\hbox{$>$\kern-.75em\lower1ex\hbox{$\sim$}}}}

\begin{document}
 \ifpdf
\DeclareGraphicsExtensions{.pdf,.jpg,.mps,.png,.eps}
 \else
\DeclareGraphicsExtensions{.eps,.ps}
 \fi

\title{Cross section dependence of event rates at neutrino telescopes}
\author{S~Hussain$^{1}$, D.~Marfatia$^{2}$, D.~W.~McKay$^{2}$ and D.~Seckel$^{1}$}
\affiliation{$^1$Department of Physics and Astronomy, University of Delaware, Newark, DE 19716}
\affiliation{$^2$Department of Physics and Astronomy, University of Kansas, Lawrence, KS 66045}

\begin{abstract}

We examine the dependence of event rates at neutrino telescopes on the 
neutrino-nucleon cross section for neutrinos with energy above 1~PeV, and 
contrast the results with those for cosmic ray experiments. 
Simple scaling of the Standard Model cross sections leaves the rate of 
upward events essentially unchanged. Details, such as detector depth and 
cross section inelasticity, can influence rates. Numerical estimates of 
upward shower, muon and tau event rates in the IceCube detector confirm 
these results.


\end{abstract}

\maketitle

{\bf{Introduction.}} 
Neutrinos with energy above $\cal{O}$($10^5$) GeV have not been detected. They are 
sought after in both neutrino telescopes and cosmic ray experiments for 
two reasons. One, they may point back to (new) sources at cosmological 
distances and reveal the structure of their internal engines. Two, they 
enable the study of neutrino interactions at energies far beyond that 
achievable at colliders. The latter is especially interesting from a 
particle physics point of view because neutrinos only interact weakly
according to the Standard Model (SM). Any new physics that affects how 
particles interact is therefore more likely to be detected in the 
neutrino sector.  However, practical issues hamper the 
extraction of such information. Knowledge of astrophysical sources and 
fundamental particle interactions must be built up simultaneously, since 
we are largely ignorant of both in the energy regime of interest, and 
knowledge of one is dependent on knowledge of the other.

Currently, the situation is further complicated by confusion. Cosmic ray 
experiments detect upcoming leptons produced in the Earth by charged 
current interactions of ``Earth-skimming" neutrinos. The effective area 
for detecting these upward leptons is a narrow projection of the actual 
detector surface area. The cross section dependence for the rate of such 
events has been derived in Refs.~\cite{kw, ffwy, afgs1}. Subsequently, 
these results have been employed to assess the potential of neutrino 
telescopes~\cite{afg,icehep}, where they are inapplicable. Our motivation 
is to elucidate clearly (via qualitative reasoning, analytical 
calculation and numerical analysis) how event rates at neutrino 
telescopes depend on the cross section and to show why the event rates in 
cosmic ray experiments have a different cross section dependence. 
We emphasize 
the role of the different cross sections and inelasticity in neutrino 
propagation and in the generation of upward shower and upward lepton 
events in providing probes for new physics.
We begin by using simple models to determine the general dependence of 
the event rates on the cross sections that govern propagation and 
detection, and apply these concepts to different experimental 
configurations. We then reproduce our main results analytically. Finally, 
we perform numerical calculations of the dependence of rates on 
cross section for 
the IceCube experiment~\cite{ice04}, which confirm the conclusions of the simplified models.

\textbf{Qualitative, geometric arguments.} 
Event rates are given by $\Gamma(E_\nu) = \frac{d\gamma}{dE_\nu} = 
\int d\Omega 
A_{e} \frac{d\phi}{d\Omega}$, where $A_e$ is the effective area for 
detection, and we use a diffuse, isotropic flux for 
$\frac{d\phi}{d\Omega}$. We assume that the dimensions of the detection 
volume are small compared to the length scale for neutrinos to be 
absorbed. One may then approximate $A_e = n \sigma_i V_e$, where $V_e$ is 
the effective volume over which an interaction may be detected.  $V_e$ 
can be expressed as $V_e= A_p \ell$, where $A_p$ is the area of the 
detector projected against the neutrino direction, and $\ell$ is the 
portion of the neutrino path to which the detector is sensitive; $n$ is 
the nucleon density within $V_e$; and $\sigma_i$ is the total (t), 
charged current (c) or neutral current (n) cross section corresponding to 
the process which makes the observable events.   For shower detection, 
$\sigma_i = \sigma_t = \sigma_c + \sigma_n$ as both charged and neutral 
current interactions produce showers, whereas for charged leptons $\sigma_i = 
\sigma_c$. Introducing the interaction length $\lambda_i = 1/(n 
\sigma_i)$, the effective area becomes \mbox{$A_e = A_p \ell/\lambda_i = 
A_p\ell n\sigma_i$}. 

We first consider showers from downward neutrinos. Since 
showers are contained, $\ell$ is given by the detector scale $s$, so 
$A_e = A_p s/\lambda_t$. Typically, the detector is at a shallow depth 
$d$ compared to the absorption length for neutrinos, so one may integrate 
$d\Omega$ over the whole sky. The  event rate for downward showers 
is
\begin{equation}
\Gamma_{\rm d,sh}
 = 2\pi A_p  \frac{d\phi}{d\Omega} \frac{s}{\lambda_t} \sim \sigma_t\,,
\label{eq:gamdsh}
\end{equation}
which increases with $\sigma_t$.
The second case is the production of downward leptons. The effective area 
is given by $A_e = A_p \ell/\lambda_c$. Here, $\ell$ is
determined by the lepton stopping ($dE/dx$) or decay, by the chord length 
to the surface, and by the detector size. We define 
leptons as detectable until they have lost $1/e$ of their initial energy, 
after which they would be difficult to distinguish from more numerous 
lower energy leptons. This is a simple criterion, which can 
only be refined with a full experimental Monte Carlo.
Taking $dE/dx = -\gamma - \beta E$, the stopping range for lepton $j$ is 
$1/\beta_j$ and $\gamma$ characterizes the ionization process. 
The decay length is $l_j= c \tau_j E/m_j$, and we denote the 
path length from the detector to the surface by $l(\theta)$. Qualitatively, 
$\ell = s/2 +{\rm Min}(1/\beta_j, l_j, l(\theta))$, and the event rate is
approximately
\begin{equation}
\Gamma_{\rm d,lep} \simeq  \pi A_p \frac{d\phi}{d\Omega} 
\frac{s}{\lambda_c} \sim \sigma_c\,,
\label{eq:gamdlep}
\end{equation}
which scales as $\sigma_c$. For simplicity,
we have suppressed the dependence on the depth of the detector.

Next we turn to upward events, and penetration through the Earth. We 
introduce the concept of neutrino attenuation length, $\lambda_a$, and 
attenuation cross section $\sigma_a = \sum_i \sigma_i y_i$ which is 
weighted by inelasticity. For neutral currents the inelasticity 
in the energy range under consideration, 
$\vev{y} \simeq 0.2-0.25$~\cite{gqrs}, 
accounts for 
downscattering. For $\nu_e$ and $\nu_\mu$ charged current interactions 
the charged leptons stop, so $y=1$ is appropriate. For $\nu_\tau$ the 
interplay between $\tau$ energy loss and decay gives an effective 
inelasticity in regeneration which increases with energy from about 0.6 
to 1~\cite{renoh}. For the SM at 10 PeV, $\sigma_a \simeq (0.77,~0.77,~ 
0.5) \sigma_t$ for ($\nu_e$, $\nu_\mu$, $\nu_\tau$).

It simplifies the discussion to consider the case of a shallow 
detector near the surface for which $d=0$. 
The length of the chord traversed by an upward 
neutrino is \mbox{$l = 2R \sin\theta$}, where $R$ is the radius of Earth 
and $\theta$ is the entry angle measured from the horizon. 
The attenuation length $\lambda_a$ 
limits the maximum chord length, so that the solid angle over which 
neutrinos are accepted scales as $\Omega = 2\pi \sin\theta = \pi 
\lambda_a/R$. Using \mbox{$A_e = A_p s/\lambda_t$}, the rate for upward shower 
events is
\begin{equation}
\Gamma_{\rm u,sh} = \pi A_p \frac{d\phi}{d\Omega} \frac{s}{R} 
\frac{\lambda_a}{\lambda_t} \sim \frac{\sigma_t}{\sigma_a}\,.
\label{eq:gamush}
\end{equation}
Although the event rate will depend on details of $\sigma_a$ and $\sigma_t$, 
to the extent that all cross sections scale in a similar manner, 
the upward shower rate is independent of new physics. This is a 
consequence of the competition between two opposing effects: as the cross 
sections increase the number of neutrinos entering the detector from 
below decreases with increased attenuation, but the probability for those 
neutrinos to interact increases.

The fourth case is upward leptons. We use \mbox{$A_e = A_p \ell/\lambda_c$}.
Folding in the reduced solid angle, 
\begin{equation}
\Gamma_{\rm u,lep} = \pi A_p \frac{d\phi}{d\Omega} \frac{\ell}{R} 
\frac{\lambda_a}{\lambda_c} \sim \frac{\sigma_c}{\sigma_a}\,.
\label{eq:gamulep}
\end{equation}
%
%
If neutral current and charged current cross sections change 
proportionally, then the upward lepton rate is independent of 
cross section. However, if the new physics only adds to the shower rate, 
then the rate decreases due to the decrease in acceptance 
without a compensatory increase in the charged current interaction 
probability.

Since it was one of our motivating concerns, we examine the rate for 
cosmic ray air shower experiments to detect upward going leptons. In this 
case the detection criterion is that the leptons pass through a 
horizontal planar area $A_0 \sim s^2$ corresponding to the footprint of 
the cosmic ray experiment, and the projected area is $A_p = A_0 
\sin\theta$. Since \mbox{$l = 2R \sin\theta < \lambda_a$}, $\sin\theta < \lambda_a/(2R)$, so the projected area decreases with an additional factor of 
$1/\sigma_a$. Integrating over solid angle,
\begin{equation}
 \Gamma_{\rm plane} \simeq \frac{\pi}{2} A_0  {\frac{d\phi}{d\Omega}}  
 \frac{\ell}{R} \frac{\lambda_a}{R} \frac{\lambda_a}{\lambda_c} 
     \sim \frac{\sigma_c}{\sigma_a^2}\,.
\label{eq:plane}
\end{equation}
We thus come to the conclusion of Ref.~\cite{kw} that the event rate for 
upward leptons scales as $1/\sigma_a^{2}$, where one factor of $\sigma_a$ 
comes from geometry, and one factor from $\sigma_c/\sigma_a$ assuming 
that the new physics does not increase the charged lepton production. 
However, we also see that this result does not apply to an in-ice 
detector such as IceCube, where the event rate for shower and charged 
lepton detection scales as $\sigma_t/\sigma_a$ and $\sigma_c/\sigma_a$ 
respectively.

An oversimplification that captures the essential point of the above analysis
for upward leptons is as follows. Suppose a detector can be described as a box with height $h$ and
side $s$. The horizontal surface area is $A_0=s^2$ and the effective geometric
area for incident direction $\theta$ is 
$A_p=sh\cos\theta+A_0 \sin\theta \simeq s(h+s\theta)$. A detector like IceCube
has $h\simeq s$ and $A_p\simeq A_0$ which is independent of $\theta$. 
Cosmic ray
detectors have a planar geometry for which $h \ll s\theta$ and hence 
$A_p\simeq A_0\theta$. 
It is this additional factor of $\theta$ that yields the 
extra factor of $1/\sigma_a$ in Eq.~(\ref{eq:plane}).


%
%

\textbf{Analytic model.} 
The probability that an upcoming neutrino propagates a distance $x$ 
along a chord is \mbox{$P(x)=e^{-x/\lambda_a}$}, and the probability that it 
interacts and produces a detectable signature $i$ in the interval $dx$ is 
$dx/\lambda_i$. The rate of interactions is
%
\mbox{$\Gamma =A_p{\frac{d\phi}{d\Omega}}
\int_{\ell}^{2R} 2\pi\frac{dl}{2R} \int_{l -\ell}^{l}
e^{-x/\lambda_a}\frac{dx}{\lambda_i}$},
%
which on integration gives
$\Gamma =\pi A_p{\frac{d\phi}{d\Omega}}\frac{\lambda _a^{2}}{R \lambda_i}
(1-e^{-{\ell}/{\lambda_a}})(1-e^{-(2R-\ell)/\lambda_a})$.
In the limit that $2R \gg \lambda _a \gg \ell$, 
$\Gamma \simeq  \pi A_p {\frac{d\phi}{d\Omega}}\frac{\ell}{R}
\frac{\lambda_a}{\lambda_i}$,
which agrees with Eqs.~(\ref{eq:gamush}) and (\ref{eq:gamulep}).

For a surface detector an additional factor of 
$\sin\theta = l/(2R)$ must be included in the $dl$ integral 
to project out the 
normal component of the lepton flux emerging from the Earth. 
For $2R \gg \lambda_a \gg \ell$, we recover Eq.~(\ref{eq:plane}).

\textbf{Numerical results.}  
So far, we have discussed neutrino event rates in various settings, but with 
several simplifying assumptions. We have used a shallow detector, 
a constant density Earth model, and have simplified the treatment of
regenerated and downscattered neutrinos. To test that 
our conclusions are robust, we perform numerical integrations to 
estimate the event rates for upward showers and leptons produced by an 
isotropic total flux of neutrinos and antineutrinos 
$\frac{d\phi}{d\Omega} = 6 \cdot 10^{-8} 
(E_\nu/{\rm GeV})^{-2}$~(cm$^2$.s.sr.GeV)$^{-1}$~\cite{wb} 
with an assumed flavor
ratio at Earth of 1:1:1. The initial neutrino energies extend from $10^6$
to $10^{12}$ GeV.
This model, used in Refs.~\cite{afg,icehep}, is for illustration only.  
A well-motivated alternative model produces a considerably larger flux 
relevant to IceCube~\cite{ahlers}. 

We numerically solve the coupled Boltzmann equations~\cite{hm} for three 
neutrino species and $\tau$ leptons, using a realistic Earth 
model to determine the flux of neutrinos at the detector. We 
include downscattering, $dE/dx$ for $\tau$'s, and neutrino production in 
$\tau$ decay. For numerical integration we divide the flux into 120 
logarithmic energy bins. We solve for the flux in each of 41 angular 
bins, logarithmically spanning a range of $\theta$ from $\sim 0.1$ to 90 
degree. The angular binning allows us to focus on narrowing angular 
ranges near the horizon as the cross sections increase. 

Event rates are determined using ice as the detector medium for both 
showers and propagating charged leptons. We do not attempt to distinguish 
different event topologies, \eg double bang, lollipop. Rather, we keep 
the simple criteria of the previous sections, \ie that all showers are 
detected and that leptons have a practical range $\ell = s/2 + {\rm 
Min}(1/\beta_j, l_j)$. 
We use $A_p = 1.4~{\rm km}^2$ and $V_e = 2.25~{\rm 
km}^3$, IceCube values~\cite{icehep} appropriate to the middle of 
the energy range under consideration, 
and infer $s=1.6~{\rm km}$. We use $\beta_\mu = 3.9 
\cdot 10^{-6}$~cm$^{-1}$ and $\beta_\tau = 0.78 \cdot 10^{-6}$~cm$^{-1}$, 
average values for ice in the same energy range.

We integrate across angles and present rates in logarithmic energy bins 
from $E_\nu= 10^{6.5}-10^{9}$~GeV. The 
$\Delta \log_{10}E_\nu=0.5$ bin width 
is motivated by 
the eventual need to separate a smoothly changing flux and SM 
cross section from a hypothetical new physics cross section which may 
`turn on' over approximately a decade in energy. Our energy bins are for 
neutrino energy at the detector. A given bin may include down scattered 
neutrinos, or neutrinos produced in propagation by $\tau$ decay. 


We parameterize physics models by three numbers $\alpha_c$, $\alpha_n$, 
and $\alpha'_n$. The first two reflect the strength of charged and 
neutral current reactions, $\sigma_i = \alpha_i \sigma_t^{\rm SM}$, 
normalized to the SM {\em total} cross section. For our 
calculations, we take the inelasticity $d\sigma_i/dy$ to be the same as 
in the SM, and assume that the $\alpha_i$ are independent of 
energy. For $\alpha'_n$ we define a new neutral current process also 
normalized to $\sigma_t^{\rm SM}$, but which is completely inelastic.
Thus, the SM would be 
described by $(\alpha_c, \alpha_n, \alpha'_n) = (r_c,r_n,0)$, where $r_i 
= \sigma_i^{\rm SM}/\sigma_t^{\rm SM}$.

\begin{figure}[ht]
\vskip 0.2in
\mbox{\includegraphics[angle=-90,width=3in]{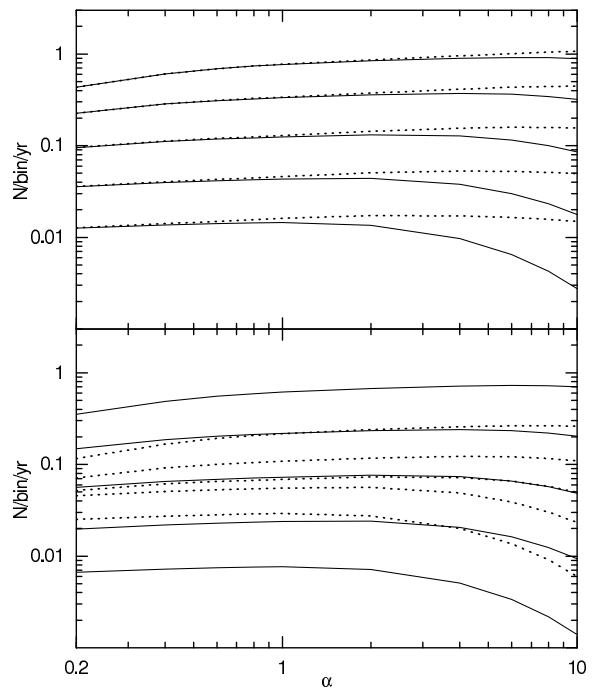}}
\caption[]{Expected event rates at IceCube for $10^{6.5} <E_\nu 
<10^9$~GeV (in five bins of width $\Delta\log_{10}E_\nu=0.5$)  
for variation in strength of SM cross sections. Top: showers, 
solid - 1.9 km deep detector, dot - shallow detector. 
Bottom: leptons, solid - $\mu$, dot - 
$\tau$. In each case, the uppermost curve corresponds to 
the lowest energy bin. We have adopted the flux of Ref.~\cite{wb}.
\label{fig:sm}}
\end{figure}

In Fig.~\ref{fig:sm}, we show the expected number of IceCube events per year as a 
function of $\alpha$ for case I: a model with $(\alpha_c, \alpha_n, 
\alpha'_n) = (\alpha r_c, \alpha r_n, 0)$, such as might occur if QCD 
saturation effects alter the growth predicted for neutrino 
nucleon cross sections~\cite{glr}. The upper panel shows shower rates in 
the five energy bins as a function of cross section for the nominal 
IceCube depth of 1.9~km. For moderate values of $\alpha$ the rates 
are essentially independent of cross section, confirming Eq.~(\ref{eq:gamush}).
 For low values of $\alpha$ the rates decrease because 
the Earth is no longer opaque, and at some point there is no compensating 
increase in flux to account for the reduced cross section. For large 
values of $\alpha$ the rates decrease because the horizontal distance
to the surface, $\sqrt{2 d R}$,
becomes comparable to the attenuation length $\lambda_a$. To confirm 
this, we redid the calculations, except now with a shallow detector 
($d=0$) just below the surface. The results are consistent with a flat 
event rate, even for large $\alpha$. The slight increase in rates at 
large $\alpha$ is due to a decreased Earth density (and increase in 
$\lambda_a$) sampled by the neutrinos as their paths are constrained to 
lie closer to the surface. Since the rates are independent of 
cross section, the integrated rate in a bin drops as $1/E_\nu$ for an 
$E^{-2}_\nu$ differential spectrum.

The lower panel shows similar results for $\mu$ and $\tau$ leptons, confirming
Eq.~(\ref{eq:gamulep}). The 
main feature to note is that at low energy the $\tau$ rates are low since 
$\tau$ decay limits $\ell$ to the detector size, while at high energies 
$\tau$ dominates since decays are delayed and $\beta_\tau < \beta_\mu$. 
Different assumptions about event identification (\eg requiring double 
bang for $\tau$) would lead to different relative rates for $\mu$ and 
$\tau$. Integrating our results above 1 PeV, we find an expectation of 
2.4, 2.2 and 0.9 events per year for upward shower, muons and taus, 
respectively. 

\begin{figure}[ht]
\mbox{\includegraphics[angle=-90,width=3in]{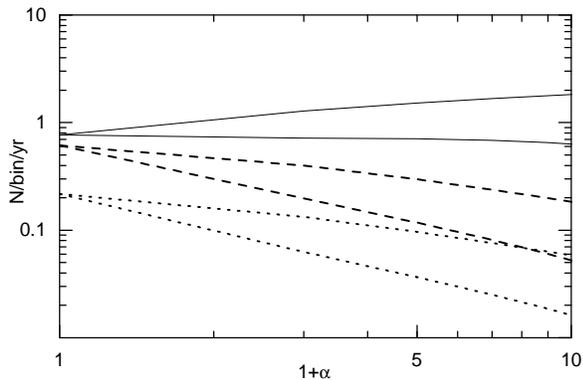}}
\caption[]{Expected event rates for $10^{6.5}<E_\nu<10^7$~GeV at IceCube for 
the case where new neutral current processes are added. For the ``black 
hole" case (lower of each pair) the shower rate (solid) is independent of cross section
while the rates for leptons (dash - $\mu$, dot - $\tau$) 
fall with attenuation. 
For the case of ``graviton exchange" (upper of each pair) reduced inelasticity and 
attenuation lead to higher rates.
\label{fig:bhnc}}
\end{figure}

In Fig.~\ref{fig:bhnc} we show results for case II, a scenario where the 
SM cross sections are unaffected, but new interactions 
produce showers, \ie $(\alpha_c, \alpha_n, \alpha'_n) = (r_c, r_n, 
\alpha)$. An example would be a low scale gravity model~\cite{add} where 
microscopic black holes are produced by neutrino-nucleon interactions and 
then decay~\cite{afgs1}. 

As for case I, Fig.~\ref{fig:bhnc} shows a shower rate that is 
essentially flat for moderate values of $\alpha$. The rate for 
leptons, however, falls inversely with 
$\alpha$ because $\sigma_c$ is fixed at the 
SM value while $\sigma_a$ is changed by new physics; see 
Eq.~(\ref{eq:gamulep}). 
The neutrino fluxes 
decrease because absorption increases, but this is not compensated by an 
increasing charged current interaction rate.

Figure~\ref{fig:bhnc} also show results for a model with $(\alpha_c, 
\alpha_n, \alpha'_n) = (r_c, r_n + \alpha, 0)$, \ie where new neutral current 
interactions are added with inelasticity as in the SM, such 
as might occur due to graviton exchange. The difference between the two 
scenarios is due to differences in downscattering. In the first case 
neutrinos are completely absorbed during propagation, while in the second 
they typically lose just 24\% of their energy (at $10^7$ GeV) per 
interaction. Define the normalized attenuation cross section as 
$w_a=\sigma_a/\sigma_t^{\rm SM}$. Then the two scenarios give $w'_a \simeq 0.77 + 
\alpha$, and $w_a \simeq 0.77 + 0.24 \alpha$. For example, consider 
$\alpha=9$ on the right hand edge of Fig.~\ref{fig:bhnc}. In this case 
$w'_a/w_a \simeq 3$, a ratio which is confirmed by the pairs of model curves 
in the figure.

\textbf{Summary and discussion.}
We have made simple scaling arguments to describe the expected detection 
rates for high energy neutrinos, with a focus on the IceCube experiment 
geometry. To validate our approach, we have performed numerical 
calculations which propagate a diffuse neutrino flux through the Earth.

Our first result is that the event rates for upgoing neutrinos in volume 
detectors are independent of the overall strength of the 
cross section. 
This conclusion is valid for over a decade change in the cross section 
normalization.

Our second conclusion is that the details matter. The event rates 
depend on a competition between neutrino attenuation and interaction 
lengths. The attenuation length depends on inelasticity $y$ as well as 
cross section, so models which alter $y$ from the SM value 
will also alter the scaling conclusion. In the extreme case of new 
totally inelastic showering interactions, the lepton rates are 
suppressed by a factor $\sim 1/(1+\alpha)$ as can be seen from 
Fig.~\ref{fig:bhnc}.

Our third conclusion is that the definition of an ``event" can lead to 
apples versus oranges comparisons. In our analysis, we required
 muons to retain most of their 
production energy, but did not require tau leptons to have an identifiable 
tag. On inclusion of tau event selection criteria and the implementation
of tau energy losses as in Ref.~\cite{dhr}, the expected SM rate for
upward going taus with energy above 
1 PeV becomes 0.3 events/year.  This is significantly smaller than the
expectation in Refs.~\cite{afg,icehep}, whose extrapolation of the 
results of Ref.~\cite{dhr} above $10^8$~GeV to values down to $10^7$~GeV 
overestimates the tau flux at the detector by an order of magnitude.



The focus of this research is to understand the possibilities for 
inferring neutrino-nucleon cross sections when the neutrino flux is 
itself unknown. By taking a ratio of two separate measurements, a diffuse 
flux can be eliminated. If the ratio differs from the SM 
prediction, then something new is going on. The converse is not true - 
varying the SM cross sections, for example, would not affect 
the ratio of upward showers to upward muons. Several remedies suggest 
themselves. a) Measure downward leptons; however, bundles of muons from 
cosmic ray air showers may provide an annoying background. b) Measure the 
lepton rate near the horizon. The angular distribution is squeezed to the 
horizon as $\alpha$ increases. c) Consider downward showers, which by our 
estimates form the majority of events. Perhaps the cleanest result would 
be a measurment of upward leptons with good angular resolution near the 
horizon, normalized to a total number of showers dominated by downward 
events. IceCube seems well designed to accomplish this, but the 
feasiblity of such methods is an experimental question, made
difficult by an unknown flux and anticipated low statistics.

{\it Acknowledgments.}
We thank L.~Anchordoqui, S.~Barwick, H. Goldberg, 
D.~Hooper 
and H.~Reno for helpful communications. This research was supported in 
part by the DOE under Grant No.~DE-FG02-04ER41308, 
by NASA under Grant No.~NAG5-5390, by the NSF under 
CAREER Award No.~PHY-0544278 and Grants No.~EPS-0236913 and 
OPP-0338219,  and by the State of Kansas through the KTEC.


\end{document}